\documentclass[aps,showpacs,footinbib,bibnotes,groupedaddress,showkeys,prl,reprint]{revtex4-1}

\usepackage{graphicx}
\usepackage{amsmath}
\usepackage{bm}
\usepackage{dcolumn}
\usepackage{amsfonts}
\usepackage{amssymb}

\begin{document}
\title{Verification of the Thomson-Onsager reciprocity relation for spin caloritronics}
\author{F. K. Dejene}
\email[E-mail to: ]{\textsl{ f.k.dejene@gmail.com}}
\author{J. Flipse}
\author{B. J. van Wees}
\affiliation{\textit{Physics of Nanodevices, Zernike Institute for Advanced Materials, University of Groningen, Groningen, The Netherlands}}
\date{\today}

\begin{abstract}
We investigate the Thomson-Onsager relation between the spin-dependent Seebeck and spin-dependent Peltier effect. To maintain identical device and measurement conditions we measure both effects in a single Ni$_{80}$Fe$_{20}$/Cu/Ni$_{80}$Fe$_{20}$ nanopillar spin valve device subjected to either an electrical or a thermal bias. In the low bias regime, we observe similar spin signals as well as background responses, as required by the Onsager reciprocity relation. However, at large biases, deviation from reciprocity occurs due to dominant nonlinear contribution of the temperature dependent transport coefficients. By systematic modeling of these nonlinear thermoelectric effects and measuring higher order thermoelectric responses for different applied biases, we identify the transition between the two regimes as the point at which Joule heating start to dominate over Peltier heating. Our results signify the importance of local equilibrium for the validity of this phenomenological reciprocity relation.
\end{abstract}
\maketitle
A linear response description of near equilibrium processes relates generalized fluxes $J_i$ to their generalized driving forces $X_{j}$ through the Onsager or transport coefficients $L_{ij}$ as $J_i$=$\sum_{j} L_{ij} X_j$ \cite{onsager_reciprocal_1931,callen_application_1948,miller_thermodynamics_1960}. The Onsager reciprocity relations (ORR) that express the coupled transport of two or more processes state that $L_{ij}$=$L_{ji}$. These symmetry relations, widely applicable in thermoelectrics \cite{onsager_reciprocal_1931,callen_application_1948}, mesoscopic charge transport studies \cite{buttiker_1986}, spintronics \cite{brataas_current_2012,Jacqoud_opnsager_2012} and spin caloritronics \cite{bauer_spincaloritronics_2010, bauer_spincaloritronics_2012}, are useful in reducing the number of independent transport coefficients \cite{xu_verification_2006} and understanding the underlying physics. In thermoelectrics, the Thomson (Kelvin) relation links the Seebeck coefficient ($S$), that describes the efficiency of thermovoltage generation in response to a temperature gradient, to the Peltier coefficient ($\Pi$), that describes the reverse process, as \cite{onsager_reciprocal_1931,callen_application_1948}
\begin{equation}
\Pi=ST_0,
\label{eq:1}
\end{equation}
where $T_0$ is the operating temperature. In linear response, the transport coefficients are assumed constant (independent of temperature) \cite{rowe_handbook_2006}. Any nonlinear contributions can lead to deviations from Eq.~\eqref{eq:1} resulting in $L_{ij}\neq L_{ji}$. Spin-dependent thermoelectric coefficients are also expected to follow this relation. Separate measurements of these coefficients in nonlocal \cite{slachter_thermally_2010,Erekhinsky_spin-dependent_2012} and pillar spin valves \cite{flipse_direct_2012,dejene_spin-dependent_2012,dejene_spin_2013}, for different measurement conditions, showed that the spin-dependent Seebeck $S_S$ and spin-dependent Peltier $\Pi_S=S_ST_0$ coefficients also obey ORR. 
 \begin{figure}[ht]
\includegraphics[width=8.6cm]{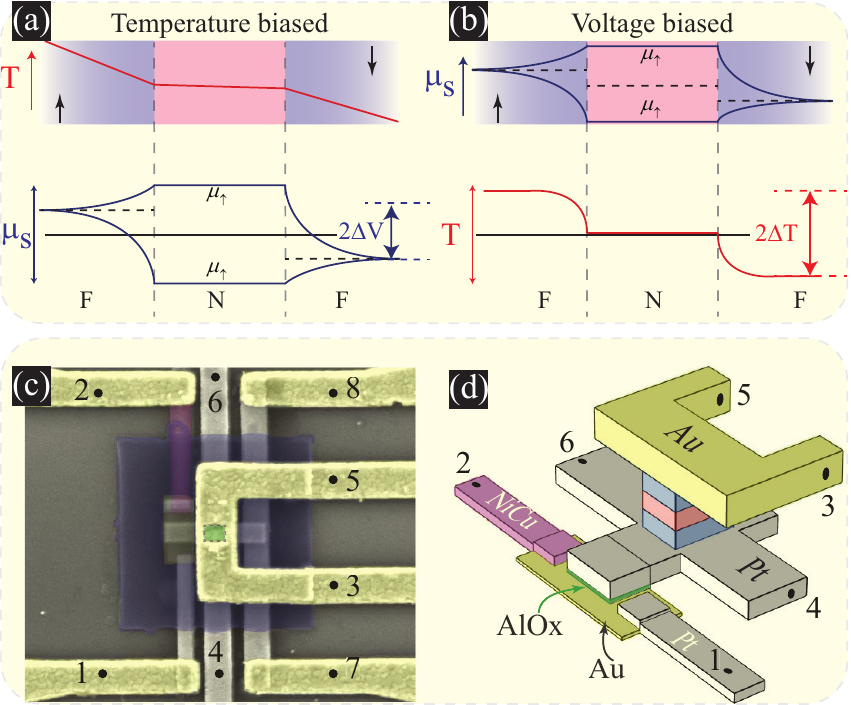}
\caption{\label{fig:fig1}(Color online) (a) SDSE in a nanopillar spin valve in the antiparallel configuration. Thermally driven spin current in the bulk of F, when injected into N, causes a spin accumulation profile shown below resulting in a voltage drop $\Delta V$ at the two F/N interfaces. (b) SDPE in a voltage-biased nanopillar spin valve. Spin accumulation in N drives a spin current that heats/cools the interface leading to a temperature change $\Delta T$ at the F/N interfaces. (c) Scanning electron microscope image and (d) Schematic diagram of the device. The electrically isolated thermocouple (contacts 1 and 2) is used to generate (detect) the temperature changes. The top Au (yellow) and bottom Pt contact (gray) sandwiches then nanopillar spin valve (green rectangle in (c)).}
\end{figure}
The formal validation of the ORR however requires that both coefficients be measured in the linear regime and more importantly in a single device \cite{miller_thermodynamics_1960}. Recent observation of the ORR for 'charge-only' thermoelectric transport in mesoscopic quantum \cite{matthews_experimental_2013} and microscopic transition ferromagnetic films \cite{avery_peltier_2013} benefited from these two strategies. 
\begin{figure*}[t]
\includegraphics[width=18cm]{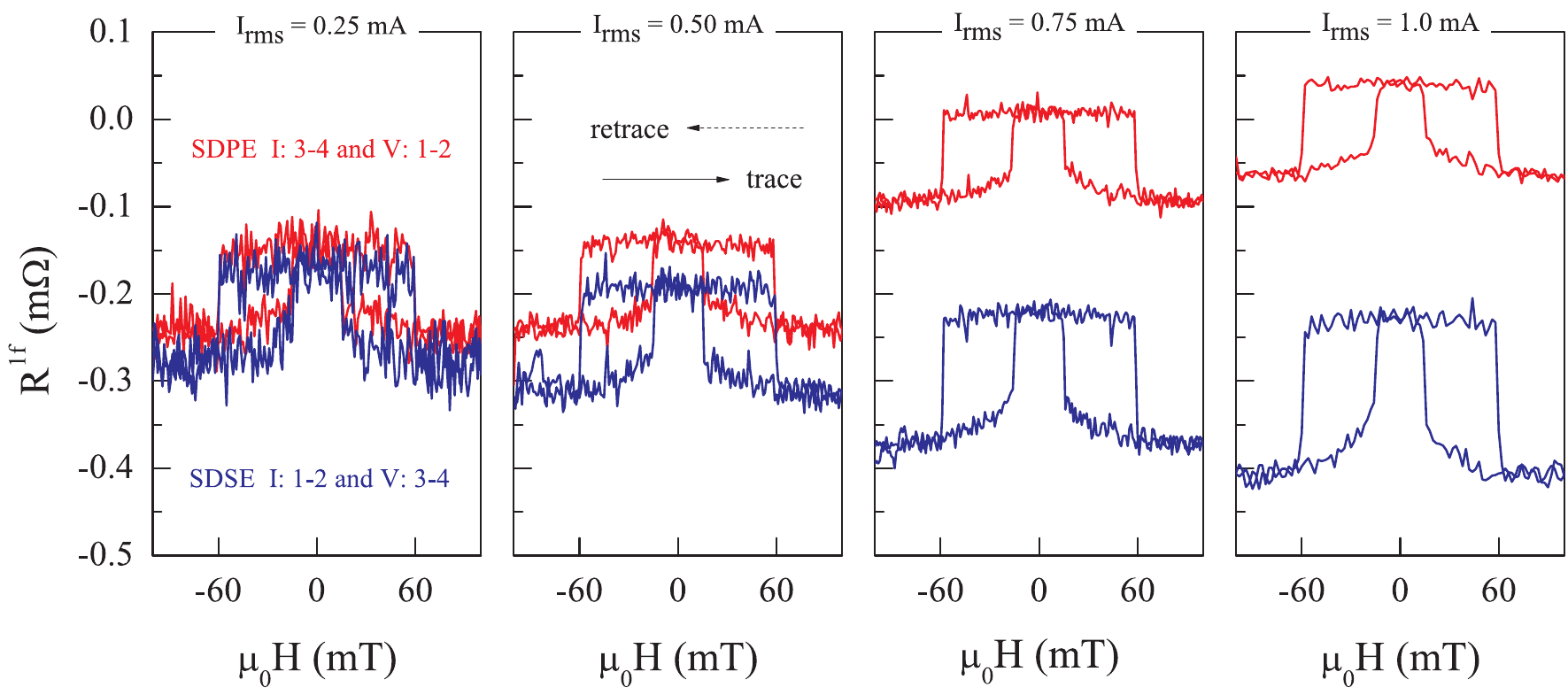}
\caption{\label{fig:fig2}(Color online)  Current dependent measurements of SDSE (blue) and SDPE (red) for an r.m.s. current of (a) 0.25 mA (b) 0.5 mA (c) 0.75 mA and (d) 1 mA. The current and voltage contacts are also shown in (a). Four-abrupt jumps in $R^{1f}$ occur when the relative magnetic configuration of the nanopillar goes from the P to AP state and back. At the low bias regime ORR is valid both for the background and spin signals. At the large bias regime deviation from ORR is observed due to nonlinear thermoelectric effects. Dipolar (magnetostatic) coupling of the two F layers favors the antiparallel state at zero field.}
\end{figure*}

In this communication, we verify the ORR between these two coefficients by measuring both the spin-dependent Seebeck effect (SDSE) and spin-dependent Peltier effects (SDPE) in a single nanopillar spin valve under identical device conditions. The device, shown in Fig.~\ref{fig:fig1}(c,d), can be subjected to either an electrical or a thermal bias. An electrically isolated thermocouple is used to generate heat (in SDSE measurement) or record temperature differences (in SDPE measurement). We find that this relation is strictly valid in the linear (low bias) regime while deviation from the ORR is observed for the nonlinear (large bias) regime. 

In the SDSE, an ac-current $I$=$I_0 \sin(2\pi f t)$ through the thermocouple (contacts 1 and 2) results in Peltier heating/cooling ($\propto I$) at the NiCu-Au and Au-Pt interfaces and Joule heating ($\propto I^2$) along the entire current path. The resulting vertical temperature bias across the nanopillar results in the injection of a spin current $j_s\propto S_S\Delta T_{pillar}$ from the ferromagnet (F) to the non-magnet (N). Here $S_S=S_{\uparrow}-S_{\downarrow}$ is the spin-dependent Seebeck coefficient in the ferromagnet \cite{hatami_thermoelectric_2009,slachter_thermally_2010,dejene_spin-dependent_2012,jansen_thermal_2012,walter_seebeck_2011} and $\Delta T_{pillar}$ is the temperature bias across the nanopillar. It is possible to modulate this spin current and the associated spin accumulation $\mu_s=\mu_{\uparrow}-\mu_{\downarrow}$ by changing the magnetic state of the nanopillar \cite{dejene_spin-dependent_2012,hatami_thermoelectric_2009}. Figure \ref{fig:fig1}(a) shows the electrochemical potential profile for spin-up and spin-down electrons \footnote{Here $P_\sigma$ is defined such that spin-up corresponds to the spin with larger electrical conductivity.} for a nanopillar spin valve subjected to a temperature bias, in the antiparallel configuration. The sum of the two voltage drops $\Delta V$ at the F/N interfaces is what is measured experimentally, using contacts 3 and 4. 

The SDPE describes the reverse process, heating/cooling of the F/N interfaces as a result of a spin current $j_S$=$j_{\uparrow}-j_{\downarrow}$ \cite{gravier_spin-dependent_2006,flipse_direct_2012} due to a gradient in $\mu_s$. In this measurement a charge current flowing through the nanopillar (using contacts 3 and 4) generates a $\mu_s$ in the N. Because $\Pi_S=0$ in N, a spin current in N does not transport heat to/away from the F/N interface. However, in F, $\Pi_s\neq 0$ and a spin current is associated with a net transport of heat depending on the magnetization of F. The resulting temperature change of $\Delta T\propto\Pi_S j_s$ at the two F/N interfaces is measured using contacts 1 and 2. Figure~\ref{fig:fig1}(b) shows this temperature profile for a nanopillar spin valve subjected to a voltage bias. 

In the experiments, we look for similar first order responses both in the SDPE and SDSE as proof for ORR. Assuming nonlinear response of up to the third order the total voltage response can be written as $V=I\cdot R_{1}+I^2\cdot R_{2}+I^3\cdot R_{3}$ where $R_i$ ($i$=1,2...) is the $i^{th}$ order response. To distinguish these various responses we employ a multiple lock-in detection technique \cite{slachter_thermally_2010,bakker_interplay_2010}. The first, second and third harmonic r.m.s. voltages measured at the lock-in amplifiers are related to $R_i$ as \cite{vera-marun_nonlinear_2012,bakker_interplay_2010}
\begin{subequations}
\begin{eqnarray}
V^{1f}&=&R_{1}I_0+\frac{3}{2}R_3I_0^3~~~\text{(in-phase)},\label{eq:2a}\\
V^{2f}&=&\frac{1}{\sqrt 2}R_2I_0^2~~~~~~~~~~\text{(90$^{o}$ out-of-phase)}~\text{and},\label{eq:2b}\\
V^{3f}&=&-\frac{1}{2}R_{3}I_0^3~~~~~~~~~~\text{(in-phase)}\label{eq:2c}.
\end{eqnarray}
\end{subequations}
In the large biasing regime, the first harmonic resistance $R^{1f}=V^{1f}/I_0$ is not the equal to the first order response $R_1$ obtained from Eq.~\eqref{eq:2a} , in which case, a correction for the contribution from the third harmonic is needed, as discussed later. All electrical measurements are performed at room temperature with slowly varying ac current such that steady state temperature distribution is reached.

Figure~\ref{fig:fig2} summarizes the main results of the paper where the first harmonic response $R^{1f}$ is plotted as a function of applied magnetic field for various values of current. The contact configurations and the root-mean-square values of the charge current used are also specified in Fig.~\ref{fig:fig2}(a). The red curves correspond to a SDPE measurement (I: 3--4 and V :1--2) and the blue curves are when the role of the current and voltage leads is reversed (I: 1--2 and V:3--4). In the SDSE, for a current of 0.25 mA through the thermocouple, we observe a spin signal $R^{1f}_P-R^{1f}_{AP}$ of $-$0.10 m$\Omega$ due to the Peltier-heating induced vertical temperature gradient across the nanopillar. In the SDPE, for a similar current through the nanopillar, the observed background and spin valve signals are identical to the ones observed in the SDSE with both measurements collapsing on each other into one indistinguishable curve within the noise level. This indicates that the SDSE voltage across the nanopillar, governed by $S_S$, is equal to the the SDPE induced thermovoltage at the thermocouple governed by $\Pi_S$. In other words, Eq.~\eqref{eq:1} is also valid for the spin-dependent counterparts of the charge Seebeck and Peltier coefficients. In the large biasing regime, say 1 mA, the spin signal of about $-0.2$ m$\Omega$ in the SDSE is twice larger than that in the SDPE. Furthermore, the background signal in the SDSE is also larger. These differences can be ascribed to deviation from the linear response regime due to higher order (nonlinear) thermoelectric effects. 
\begin{figure}[t]
\includegraphics[width=8.6cm]{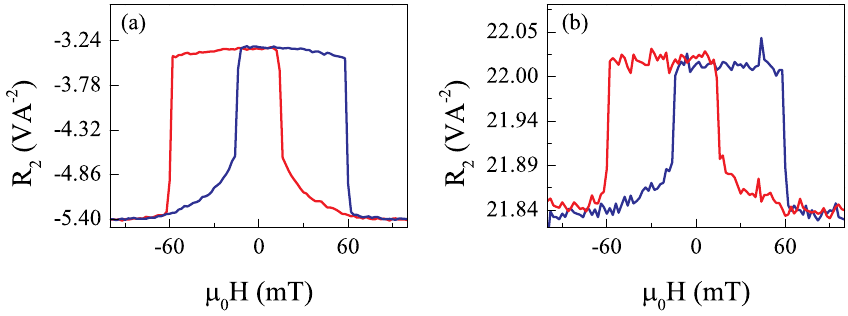}
\caption{\label{fig:fig3}(Color online) Second order response $R_{2}$ obtained from the measured $V^{2f}$ via Eq.~\eqref{eq:2b} as a function of applied magnetic field for (a) SDSE and (b) SDPE at a current of 1 mA.}
\end{figure}

\begin{figure}[t]
\includegraphics[width=8.6cm]{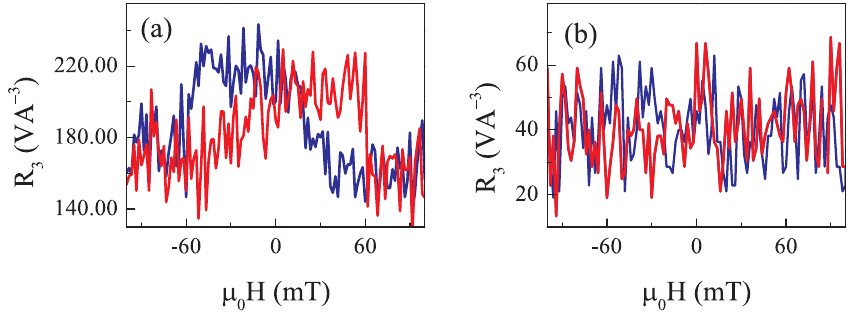}
\caption{\label{fig:fig4}(Color online) Third order response $R_{3}$ obtained from the measured $V^{3f}$ via Eq.~\eqref{eq:2c} as a function of applied magnetic field for (a) SDSE and (b) SDPE at a current of 1 mA.}
\end{figure}

In addition to the first order response due to Peltier heating, we also observe higher order responses (Fig.~\ref{fig:fig3} and \ref{fig:fig4}). The magnetic field dependence of the second order response $R_2$, for the SDSE (Fig.~\ref{fig:fig3}(a)) and SDPE (Fig.~\ref{fig:fig3}(b)), shows a spin signal $R_{2S}$ of $-$1.9 VA$^{-2}$ and $-$0.2 VA$^{-2}$, respectively. The physical origin of the spin signal in the SDSE is identical to that in Fig.~\ref{fig:fig2}, but now due to the Joule-heating induced vertical temperature gradient across the nanopillar. The spin signal of observed in the SDPE (Fig.\ref{fig:fig3}(b)) is not however related to the spin-dependent Seebeck coefficient. Rather it originates from the change in the nanopillar resistance (and associated Joule heating) when the magnetic state of the nanopillar changes from the P to AP configuration \cite{flipse_direct_2012}. 

In the large biasing regime, a spin signal is also observed in the third order response $R_3$ of the SDSE measurement (Fig.~\ref{fig:fig4}(a)) while no spin signal (above the noise level) is present in the SDPE (Fig.~\ref{fig:fig4}(b)). This observation, that points to the presence of nonlinear thermoelectric effects, is consistent with the nonlinear bias- dependence observed in Fig.~\ref{fig:fig2}. From Eq.~\eqref{eq:3} it becomes clear that the combined effect of Joule and Peltier heating or concurrent changes in the the material properties of both the nanopillar and thermocouple can lead to the third order response \cite{bakker_interplay_2010}. In this regime, the first harmonic voltage $V^{1f}$ is not strictly linear with the applied current and hence should be corrected for the contribution from the third harmonic response as $V_1$=$V^{1f}$+$3V^{3f}$ (see Eq.~\eqref{eq:2a}).  

\begin{figure*}[t]
\includegraphics[width=18cm]{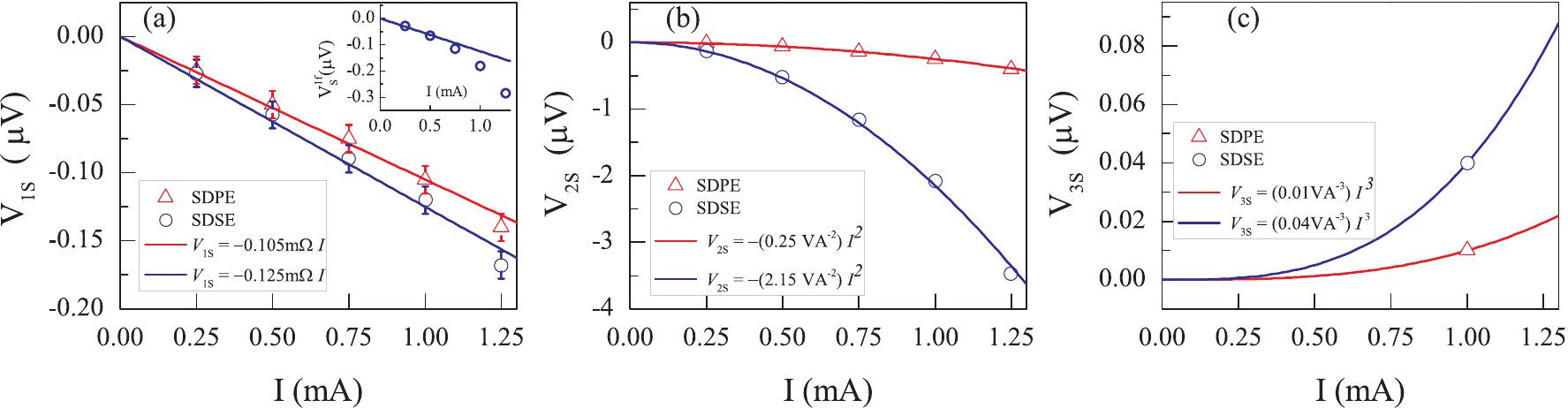}
\caption{\label{fig:fig5}(Color online)  Current dependence of the (a) $V_{1S}$ (b) $V_{2S}$ and (c) $V_{3S}$ in the SDPE (triangles) and SDSE (circles). Deviation from the expected scaling with bias (lines) occurs when Joule heating dominates over Peltier heating. The error bars in (a) indicate the maximum noise level. Inset in (a) is the spin signal as obtained directly from the measured first harmonic response in the SDSE showing non-linearity due to a contribution from $V_{3S}$ (see Eq.~\eqref{eq:2a}).}
\end{figure*}

Next, we discuss the bias dependence of the spin signals, the difference between the parallel and antiparallel voltages, for each of the first order ($V_{1S}$=$R_{1S}I$), second order ($V_{2S}$=$R_{2S}I^2$) and third order ($V_{3S}$=$R_{3S}I^3$) responses (Fig.~\ref{fig:fig5}). While the uncorrected first harmonic signal in the SDSE (shown in the inset of Fig.~\ref{fig:fig5}(a)) is rather nonlinear with applied bias, the corrected first order response (main plot of Fig.~\ref{fig:fig5}(a)) scales linearly with the applied bias, both in the SDPE (triangles) and SDSE (circles). The slopes of the linear fits are also close to each other, within 20\%, indicating validity of ORR over the entire bias range studied here. The current-dependence of the second order spin signal is also shown in Fig.~\ref{fig:fig5}(b). Absence of any deviation from the expected quadratic dependence on the applied bias supports our assumption of nonlinear response up to the third order.
Note that the current at which $V_{2S}$=$V_{1S}$ marks the point at which Joule heating is equal to Peltier heating. These current values of 50 $\mu$A (in SDSE) and $0.5$ mA (in SDPE) can be taken as threshold values beyond which nonlinear thermoelectric processes become relevant for our nanopillar spin valves, which is consistent with Fig.~\ref{fig:fig2}. 

For the sake of completeness, the bias dependence of the the third order spin signal is shown in Fig.~\ref{fig:fig5}(c). Because these higher order effects are only visible in the large biasing regime, only $V_{3S}$ data at a current of 1 mA is shown. The solid lines are cubic dependencies extrapolated to the linear regime. These third order response spin signals are subtracted from the measured first harmonic spin signal (shown in Fig.~\ref{fig:fig2}) in order to obtain the first order spin signals in Fig.~\ref{fig:fig5}(a).

To understand the deviation from ORR we look at the thermovoltage in the  $V_{SDSE}\propto S_S\Delta T_{pillar}$ when the local device temperature increases by $\delta T$=$T$-$T_0$. Noting that $S_S$ is linear with temperature as $S_{S}(T)$=$S_{S}(T_0)(1+\gamma \delta T)$ where $\gamma=1/T_0$ \cite{bakker_interplay_2010}, the nonlinear thermovoltage signal reads
\begin{equation}
V_{SDSE}\propto S_{S}(T_0)\Delta T_{pillar}+\gamma S_{S}(T_0)\Delta T_{pillar}\delta T.
\label{eq:3}
\end{equation}
When $\delta T$ and $\Delta T_{pillar}$ are a sizable fraction of $T_0$, the second term in Eq.~\eqref{eq:3} becomes important leading to a deviation from ORR. Similarly, the thermovoltage in the SDPE can be nonlinear due to the temperature dependencies of the $S$ (of the thermocouple) and $S_S$ (of the ferromagnet). 

Because it is difficult to keep track of interdependent changes in material parameters, we use a three dimensional spin dependent finite element model (3D-FEM)\cite{slachter_modeling_2011,slachter_thermally_2010,bakker_interplay_2010} to understand these nonlinear effects. The spin-dependent charge current $\vec{J}_{\uparrow,\downarrow}$ and heat current density $\vec{Q}$ are extended to include the temperature dependence of the input-material parameters as
\begin{equation}
\begin{pmatrix} \vec{J}_{\uparrow,\downarrow} \\ \vec{Q} \end{pmatrix} =- \begin{pmatrix} \sigma_{\uparrow,\downarrow}(T) & \sigma_{\uparrow,\downarrow}(T) S_{\uparrow,\downarrow}(T) \\ \sigma_{\uparrow,\downarrow}(T)\Pi_{\uparrow,\downarrow}(T) & k(T)  \end{pmatrix} \begin{pmatrix} \vec{\nabla} V_{\uparrow,\downarrow} \\ \vec{\nabla} T \end{pmatrix}
\label{eq:4}
\end{equation}
where $\sigma_{\uparrow,\downarrow}(T)$=$\sigma_{\uparrow,\downarrow}/(1$+$\alpha_T \Delta T)$ is the temperature dependent electrical conductivity, $\alpha_T$ is the temperature coefficient of resistance. The bulk values of $\alpha_T$ are well tabulated in the literature ($\sim$$10^{-3}$) and that of thin films is known to be lower than the bulk value due to, for example, enhanced electron scattering at boundaries \cite{zhang_platinum_nanofilm_2005}, which we use in our model. $\kappa(T)$ is the electronic thermal conductivity defined using the Wiedemann-Franz relation valid for metals at the temperatures of our experiments \cite{bakker_nanoscale_2012}. Following Ref.~\onlinecite{slachter_modeling_2011}, we define the spin-dependent electrical conductivity as $\sigma_{\uparrow,\downarrow}$=$\sigma(1\pm P_\sigma)/2$ where $P_\sigma$$=$$(\sigma_\uparrow$-$\sigma_\downarrow)/(\sigma_\uparrow$+$\sigma_\downarrow)$ is the spin polarization of the electrical conductivity. The spin-dependent Seebeck coefficient is given by $S_{\uparrow,\downarrow}$=$S-\frac{1}{2}(1\mp P_\sigma)S_S$. Material parameters for the modeling are taken from the literature \cite{bakker_nanoscale_2012,dejene_spin_2013}. 

In order to calculate the spin signals observed in Fig.~\ref{fig:fig5} we first extract $P_\sigma$ from a separate measurement of the electrical spin valve (not shown here). The spin polarization of the Seebeck coefficient $P_S$ was also obtained from a separate measurement of the SDSE based on the Pt-Joule heater (also not shown here but discussed elsewhere \cite{dejene_spin-dependent_2012,dejene_spin_2013}). Using the obtained values of $P_\sigma=0.58$ and $P_S=0.35$ we calculate the SDPE and SDSE signals using the 3D-FEM. For the SDPE, we obtain spin signals of $R_{1S}$=$-$95 $\mu\Omega$, $R_{2S}$=$-0.19$ VA$^{-2}$ and $R_{3S}$=$-6$ VA$^{-3}$ for the first, second and third order signals, respectively, in agreement with the measured values. For the SDSE, the calculated values of $R_{1S}$=$-$93 $\mu\Omega$ and $R_{2S}$=$-1.8$ VA$^{-2}$ are close to the measured values (see Fig.~\ref{fig:fig5}). The third order response $R_{3S}$=$-125$ VA$^{-3}$ in the SDSE is however three times larger. Although we do not understand this difference,  owing to the good agreement of the calculated signals with the measured values, we conclude that nonlinear thermoelectric effects, as modeled here in terms of the temperature dependence of the transport coefficients, can describe both the linear and higher order responses. 

In summary, we experimentally tested and verified the Onsager-Kelvin reciprocity relation for the spin-dependent Seebeck and Peltier coefficients and also provided the extent to which this reciprocity relation is respected. At small biases, when Joule heating is small, the Onsager reciprocity relation holds while, at large thermal/electrical biases, temperature dependence of both thermal and electrical transport coefficients drives the system into a non-linear regime where the basic assumption for ORR are not valid anymore. This deviation from ORR reciprocity is due mainly to nonlinear thermoelectric effects. It is therefore important to take nonlinear thermoelectric contributions into account in analyzing charge, spin and heat transport especially when the temperature gradient across a device is large. We also showed that higher order thermoelectric contributions, when not taken into account, could lead to apparent deviation from ORR. 

The authors thank M. de Roosz, J.G. Holstein, H. Adema and B. Wolfs for technical assistance. I.J. Vera-Marun for reading the manuscript. This work is part of the research program of the Foundation for Fundamental Research on Matter (FOM) and is supported by NanoLab NL, EU-FP7 ICT grant 257159 MACALO, EU-FET Grant InSpin 612759, and the Zernike Institute for Advanced Materials.
%
\end{document}